\documentclass[twocolumn,twocolappendix]{aastex63}
\usepackage{amsmath}
\usepackage{amssymb}
\usepackage{bm}
\usepackage{hyperref}
\hypersetup{colorlinks,breaklinks,allcolors=blue}
\usepackage[super]{nth}
\usepackage{arydshln}
\usepackage{multirow}

\defcitealias{Kegerreis+2020}{K20}

\usepackage{tikz}
\usetikzlibrary{calc,fadings,decorations.pathreplacing,arrows,angles}

\tikzset{%
	>=latex, 
	inner sep=0pt,%
	outer sep=2pt,%
	mark coordinate/.style={inner sep=0pt,outer sep=0pt,minimum size=3pt,
		fill=black,circle}%
}

\received{2020 July 8}
\revised{2020 August 19}
\accepted{2020 August 25}
\submitjournal{ApJL}

\shorttitle{Atmospheric Erosion by Giant Impacts}
\shortauthors{Kegerreis et al.}

\begin{document}

\title{\Large Atmospheric Erosion by Giant Impacts onto Terrestrial Planets:
  \\ A Scaling Law for any Speed, Angle, Mass, and Density}

	\correspondingauthor{Jacob Kegerreis}
	\email{jacob.kegerreis@durham.ac.uk}
	\author[0000-0001-5383-236X]{J. A. Kegerreis}
	\affiliation{Institute for Computational Cosmology, Durham University, Durham, DH1 3LE, UK}

	\author[0000-0001-5416-8675]{V. R. Eke}
	\affiliation{Institute for Computational Cosmology, Durham University, Durham, DH1 3LE, UK}
	\author[0000-0001-5646-120X]{D. C. Catling}
	\affiliation{Department of Earth and Space Sciences, University of Washington, Seattle, WA, USA}
  \author[0000-0002-6085-3780]{R. J. Massey}
  \affiliation{Institute for Computational Cosmology, Durham University, Durham, DH1 3LE, UK}
  \author[0000-0002-8346-0138]{L. F. A. Teodoro}
  \affiliation{BAERI/NASA Ames Research Center, Moffett Field, CA, USA}
  \affiliation{School of Physics and Astronomy, University of Glasgow, G12 8QQ, Scotland, UK}
	\author[0000-0002-2462-4358]{K. J. Zahnle}
	\affiliation{NASA Ames Research Center, Moffett Field, CA, USA}  




\begin{abstract}

We present a new scaling law to predict the loss of atmosphere from planetary collisions for any speed, angle, impactor mass, target mass, and body compositions, in the regime of giant impacts onto broadly terrestrial planets with relatively thin atmospheres. To this end, we examine the erosion caused by a wide range of impacts, using 3D smoothed particle hydrodynamics simulations with sufficiently high resolution to directly model the fate of low-mass atmospheres around 1\% of the target's mass. Different collision scenarios lead to extremely different behaviours and consequences for the planets. In spite of this complexity, the fraction of lost atmosphere is fitted well by a power law. Scaling is independent of the system mass for a constant impactor mass ratio. Slow atmosphere-hosting impactors can also deliver a significant mass of atmosphere, but always accompanied by larger proportions of their mantle and core. Different Moon-forming impact hypotheses suggest that around 10 to 60\% of a primordial atmosphere could have been removed directly, depending on the scenario. We find no evident departure from the scaling trends at the extremes of the parameters explored. The scaling law can be incorporated readily into models of planet formation. 

\end{abstract}

\keywords{
  Impact phenomena (779);
  Planetary atmospheres (1244); 
  Earth atmosphere (437);
  Hydrodynamical simulations (767).
}


\section{Introduction}
\label{sec:introduction}

Terrestrial planets are thought to form from tens of roughly Mars-sized embryos
that crash into each other after accreting from a protoplanetary disk
\citep{Chambers2001}.
At the same time, planets grow their atmospheres
by accreting gas from their surrounding nebula,
degassing impacting volatiles directly into the atmosphere,
and by outgassing volatiles from their interior \citep{Massol+2016}.

For a young atmosphere to survive
it must withstand radiation pressure of its host star,
frequent impacts of small and medium impactors,
and typically at least one late giant impact
that might remove an entire atmosphere in a single blow
\citep{Schlichting+Mukhopadhyay2018}.

The rapidly growing population of observed exoplanets 
reveals a remarkable diversity of atmospheres,
even between otherwise similar planets in the same system
\citep{Lopez+Fortney2014,Liu+2015b,Ogihara+Hori2020},
and the Earth's own atmosphere shows a complex history of fractionation and loss 
\citep{Tucker+Mukhopadhyay2014,Sakuraba+2019,Zahnle+2019}.
However, the full extent of the role played by giant impacts is uncertain,
in part due to the lack of comprehensive models 
for the atmospheric erosion caused across the vast parameter space
of possible impact scenarios.

A challenge for numerical simulations 
is the low density of an atmosphere compared with the planet, 
which requires high resolution \citep{Kegerreis+2019}.
For this reason, previous studies have made progress by focusing primarily
on 1D models or thick atmospheres ($>$$\sim$5\% of the total mass),
often also limited to only head-on impacts 
or too few scenarios to make broad scaling predictions
\citep{Genda+Abe2005,Inamdar+Schlichting2015,Hwang+2018,Lammer+2020,Denman+2020}.

\citet[][hereafter \citetalias{Kegerreis+2020}]{Kegerreis+2020}
used high-resolution smoothed particle hydrodynamics (SPH) simulations 
of giant impacts to investigate the detailed dependence of atmospheric loss 
on the speed and angle of an impact
and to examine the different mechanisms 
by which thin atmospheres could be eroded.
They derived a scaling law to predict the loss from any such collision
between an impactor and target similar to those of the 
canonical Moon-forming impact.
However, the study was limited to those single target and impactor masses.

We now simulate a wide range of target and impactor masses and compositions
in addition to different angles and speeds,
in order to develop a scaling law that can apply to any giant impact 
in the broad regime of terrestrial planets with thin atmospheres.
The tested scenarios include masses ranging from 
roughly three times the Earth's mass down to a few percent of its mass;
differentiated and undifferentiated planets
with densities from about half to over double the Earth's density;
angles from head-on to highly grazing; 
and speeds from 1 to 3 times the mutual escape speed.

\begin{figure}[t]
  \centering
  %
  %
  \includegraphics[width=\columnwidth, trim={11mm 251mm 118mm 12mm}, clip]{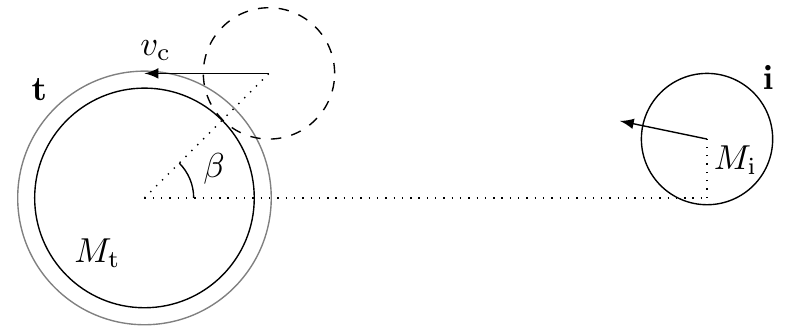}
  \caption{The initial conditions for an impact scenario, with the target
  (t) on the left and the impactor (i) on the right
  with masses $M_{\rm t,i}$, 
  shown in the target's rest frame.
  The speed and angle at first contact, $v_{\rm c}$ and $\beta$, 
  and the dimensionless impact parameter $b \equiv \sin(\beta)$
  are set ignoring the atmosphere
  and neglecting any tidal distortion before the collision.
  The initial separation is set such that the time to impact 
  under the same assumptions is 1~hour, 
  using the equations in Appx. A of \citet{Kegerreis+2020}.
  \label{fig:impact_scenario}}
\end{figure}

\begin{figure*}[t]
	\centering
	\includegraphics[width=1\textwidth, trim={76mm 29.5mm 96mm 9.5mm}, clip]{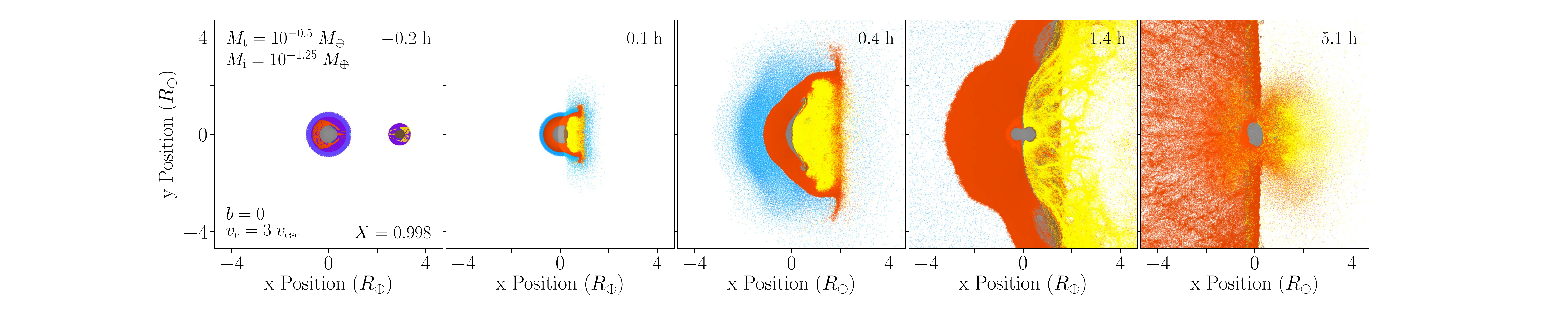}\\\vspace{-0.6mm}
	\includegraphics[width=1\textwidth, trim={76mm 29.5mm 96mm 9.5mm}, clip]{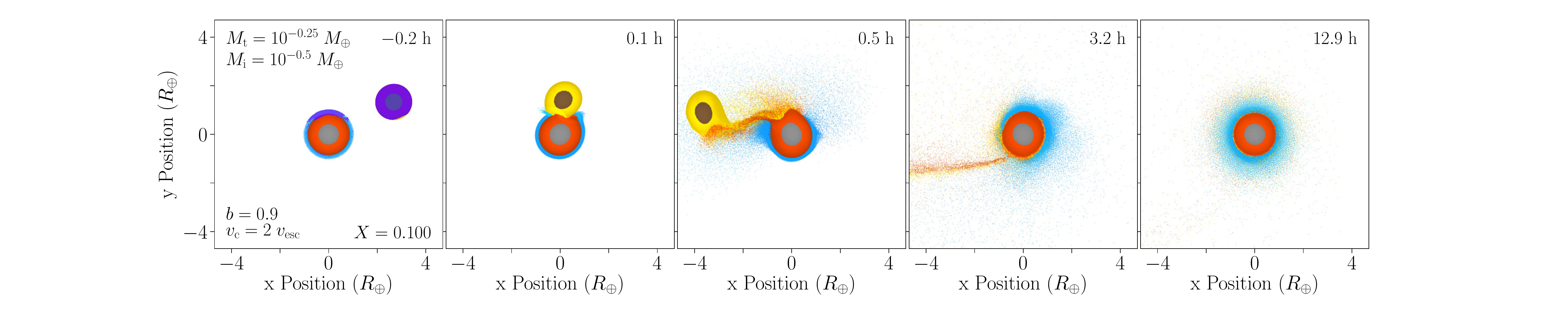}\\\vspace{-0.6mm}
	\includegraphics[width=1\textwidth, trim={76mm 29.5mm 96mm 9.5mm}, clip]{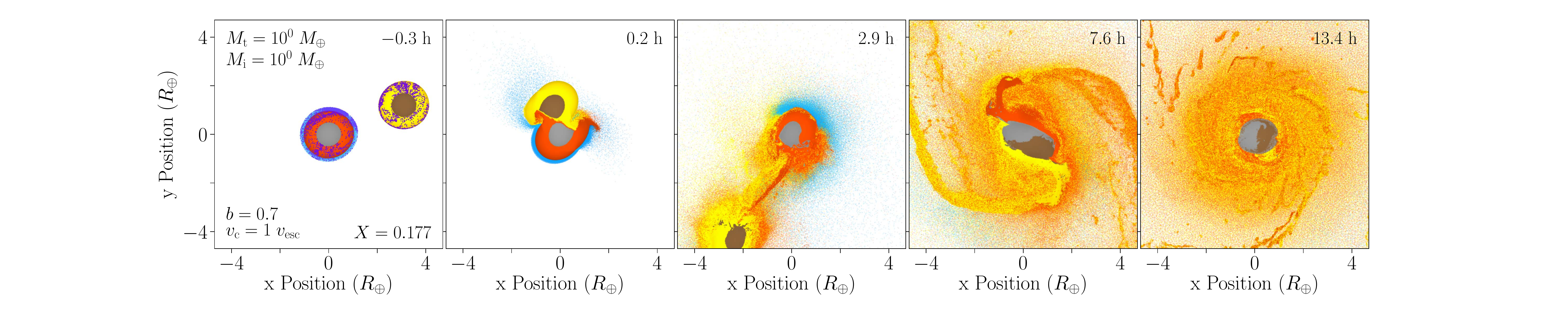}\\\vspace{-0.6mm}
	\includegraphics[width=1\textwidth, trim={76mm 8mm 96mm 9.5mm}, clip]{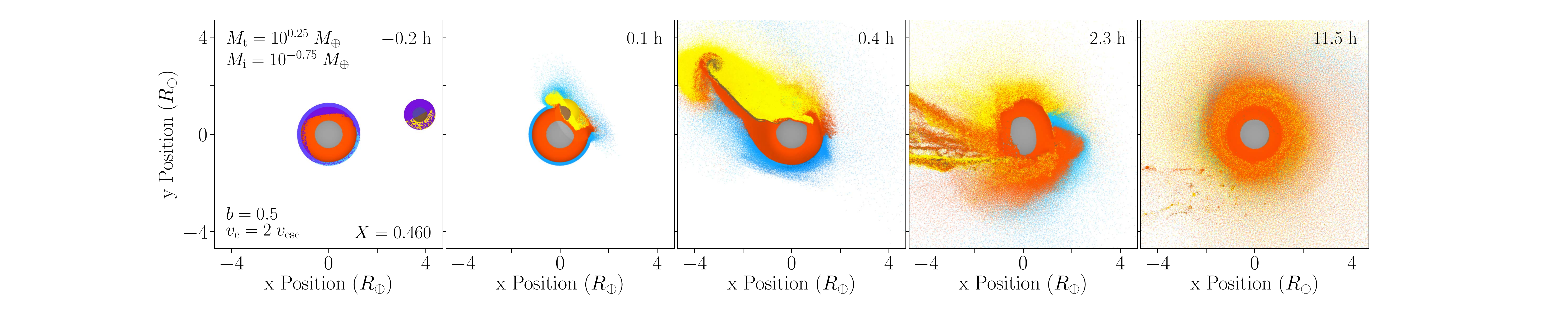}\\
  \caption{
    Illustrative snapshot cross-sections 
    from four example impact simulations,
    using $\sim$$10^{7.5}$ SPH particles.
    The annotations detail the parameters for each scenario
    (see \S\ref{sec:methods} and Fig.~\ref{fig:impact_scenario});
    the lost mass fraction of the atmosphere, $X$; and the time.
    Note that the snapshots are at different times
    to show the evolution in each case.
    In the leftmost panels, 
    the particles that will become unbound and escape the system
    are highlighted in purple on a pre-impact snapshot,
    though note that less material may be lost away from the impact plane.
    Grey and orange show the target's core and mantle material respectively,
    and brown and yellow show the same for the impactor.
    Blue is the target's atmosphere.
    The colour luminosity varies slightly with the internal energy.
    The animations (see online version or 
    \href{http://icc.dur.ac.uk/giant_impacts/atmos_fid_1e8_u_anim.mp4}{icc.dur.ac.uk/giant\_impacts})
    show the early stages of other representative impacts 
    with the particles coloured by their internal energy.
		\label{fig:demo_snaps}}
\end{figure*}

\section{Methods} \label{sec:methods}

The 259 simulations in this study can be summarised as three related suites: 
{\bf(1)} A set of impacts with different target and impactor masses, 
for head-on and grazing, slow and fast collisions. 
This also includes some scenarios with atmosphere-hosting impactors.
{\bf(2)} A set of changing-angle and changing-speed scenarios 
for a subset of target and impactor combinations. 
{\bf(3)} A set of targets and impactors with 
extreme compositions and densities,
including different equations of state. 
The full details of each suite 
and the SPH simulations 
are described in Appx.~\ref{sec:appx:methods}.
The parameters for each simulation 
including the resulting atmospheric erosion
are also listed in Table~\ref{tab:X_results}.

We specify each impact scenario by the masses of the target and impactor,
$M_{\rm t}$ and $M_{\rm i}$, 
in addition to the impact parameter, $b$,
and the speed at first contact, $v_{\rm c}$, of the impactor 
with the target's surface, as illustrated in Fig.~\ref{fig:impact_scenario}.
The radii, $R_{\rm t, i}$, are defined at the base of any atmosphere.
The speed at contact is set in units of 
the mutual escape speed of the system, 
$v_{\rm esc} = \sqrt{2 G \left(M_{\rm t} + M_{\rm i}\right) / 
\left(R_t + R_i\right)}$.
For clarity, when used to identify scenarios in the text 
the $M_{\rm t}$ and $M_{\rm i}$ labels do not include any atmosphere.

Our targets and impactors are differentiated into 
a rocky mantle and an iron core
containing 70\% and 30\% of the mass, respectively,
using the \citet{Tillotson1962} granite 
or, for a subset of 21 simulations, ANEOS forsterite \citep{Stewart+2019}
and the Tillotson iron \citep{Melosh1989} equations of state (EoS).
We also use some undifferentiated bodies made of only iron or rock.

All targets and some impactors have an added atmosphere 
with 1\% of their core+mantle mass,
using \citet{Hubbard+MacFarlane1980}'s hydrogen--helium EoS.
The planetary profiles are generated by integrating inwards 
while maintaining hydrostatic equilibrium%
\footnote{
  The WoMa code for producing spherical and spinning planetary profiles and 
  initial conditions is publicly available with documentation and examples at
  \href{https://github.com/srbonilla/WoMa}{github.com/srbonilla/WoMa},
  and the python module \texttt{woma} can be installed directly with
  \href{https://pypi.org/project/woma/}{pip}
  \citep{Ruiz-Bonilla+2020}.
},
then the particles are placed to precisely match the resulting density profiles
using the stretched equal-area (SEA%
\footnote{
  The SEAGen code is publicly available at
  \href{https://github.com/jkeger/seagen}{github.com/jkeger/seagen}
  and the python module \texttt{seagen} can be installed directly with
  \href{https://pypi.org/project/seagen/}{pip}
  \citep{Kegerreis+2019}.
})
method,
following the same procedure detailed in \citetalias{Kegerreis+2020}~\S2.1.

The simulations are run using the 
open-source hydrodynamics and gravity code SWIFT%
\footnote{
  Version 0.8.5. SWIFT is publicly available at \href{www.swiftsim.com}{www.swiftsim.com}.
}
as described in \citet{Kegerreis+2019}.
We use around $10^{7.5}$ SPH particles for each simulation,
depending on the masses of the two bodies
(see Appx.~\ref{sec:appx:methods}).
\citetalias{Kegerreis+2020} ran convergence tests 
for the fraction of atmosphere eroded by similar impacts
and similar atmosphere mass fractions to those in this study.
Simulations using $10^7$ particles yielded results 
that agreed to within $\sim$2\% with 
ones using $10^{7.5}$ and $10^{8}$ particles,
with improved convergence for more-erosive collisions,
so our somewhat higher resolution here should be comfortably sufficient.

\citetalias{Kegerreis+2020} found that the time required 
for the amount of eroded material to settle
ranges from less than 1 hour after contact 
for high-speed and/or low-angle impacts
up to 5--10~hours for slower, grazing collisions.
Depending on the scenario 
each simulation is run for a conservative 5--14~hours after contact.

\begin{figure*}[t]
	\centering
	\includegraphics[width=\textwidth, trim={8mm 9mm 18mm 6mm}, clip]{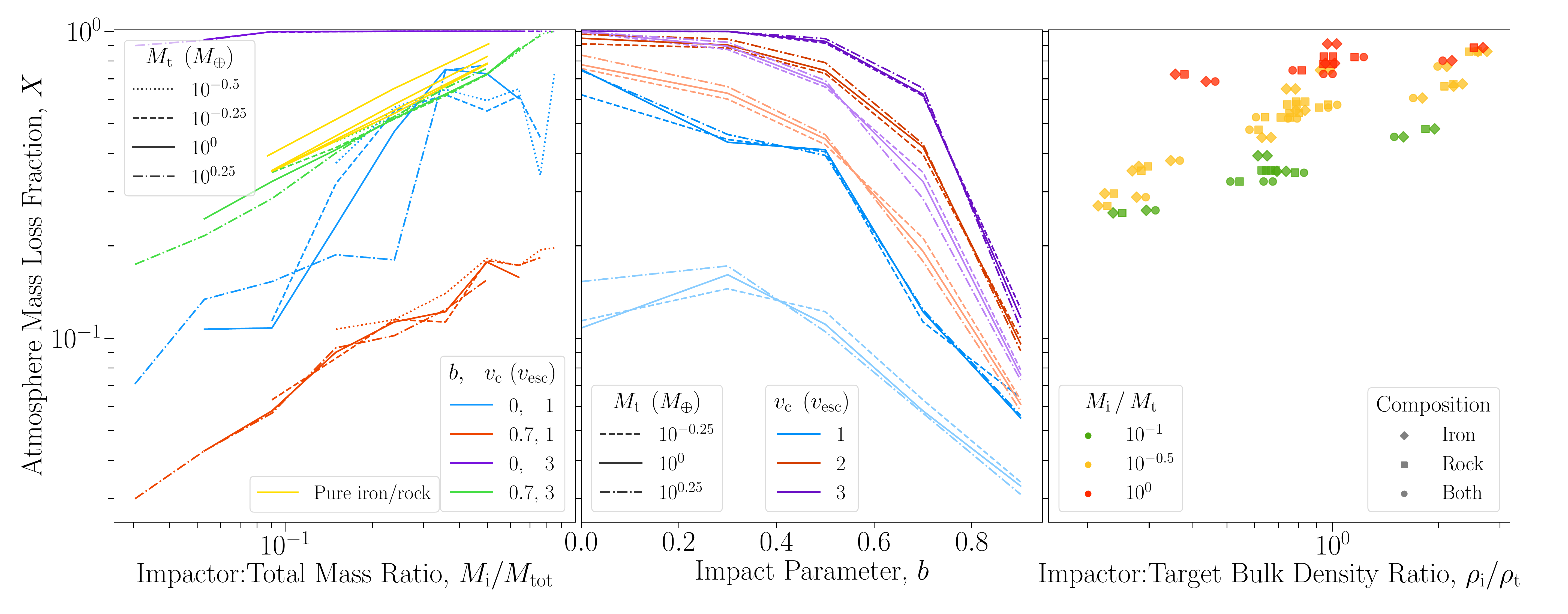}
  \caption{
    The lost mass fraction of the atmosphere as a function of:
    {\bf(left)} 
    The impactor:total mass ratio,
    plotted separately for each of the four scenarios (colours)
    and each target mass (line styles) of the first suite,
    including atmosphere-hosting impactors 
    being treated as targets to give impactor:target mass ratios 
    greater than one.
    The yellow lines show subsets of the third suite
    for pure-iron or pure-rock bodies, with $b=0.7$, $v_{\rm c}=3$.
    {\bf(middle)}
    The impact parameter,
    plotted separately for each speed at contact (colours)
    and each target mass (line styles) of the second suite.
    The subsets with an impactor:target mass ratio of $10^{-1}$ 
    are shown by the lighter colour (lower magnitude) lines
    and $10^{-0.25}$ by the darker lines, respectively.
    {\bf(right)}
    The ratio of the impactor and target's bulk densities,
    for each base impactor:target mass ratio (colours) of the third suite
    (see Appx.~\ref{sec:appx:methods}).
    The left and right markers in each pair 
    show the composition of the target and impactor, respectively,
    as detailed in the legend.
		\label{fig:X_trends}}
\end{figure*}

\begin{figure*}[t]
  \centering
  \includegraphics[width=0.95\textwidth, trim={8mm 9mm 11mm 6mm}, clip]{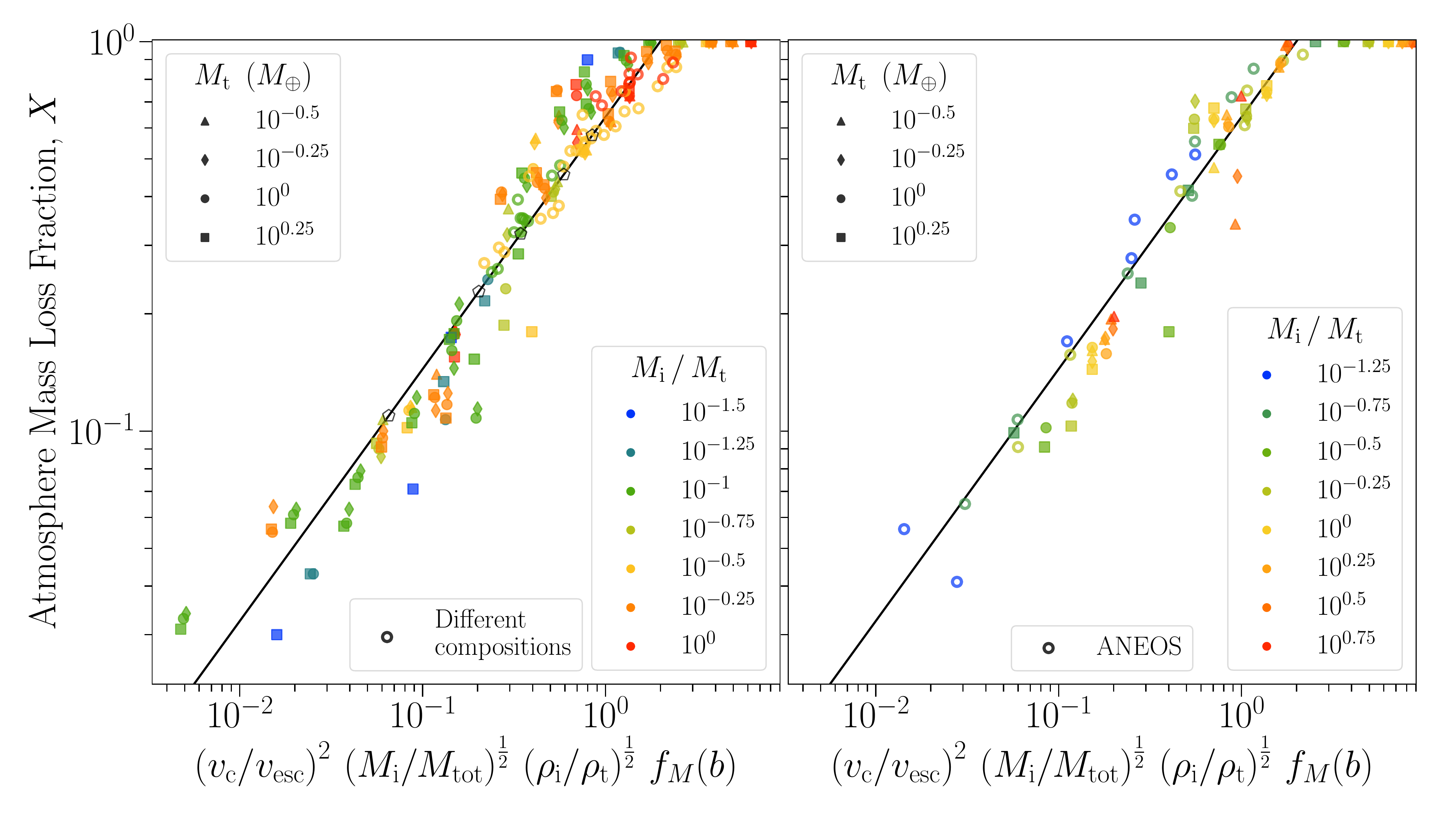}
  \caption{
  The lost mass fraction of the atmosphere:
  {\bf(left)} from all of the standard simulation scenarios
  as a function of the scaling parameter,
  coloured by the impactor:target mass ratio
  with markers set by the target mass.
  Open markers represent the third-suite impacts
  where one or both bodies are pure iron or pure rock.
  The black line shows our scaling law (Eqn.~\ref{eqn:scaling_law}).
  The black, open pentagons correspond to different Moon-forming impacts,
  as detailed in the text.
  {\bf(right)} from (1) scenarios with atmosphere-hosting impactors 
  (solid markers) -- including treating the impactor as the target
  to give impactor:target mass ratios (colours) larger than 1,
  and (2) scenarios for bodies with ANEOS forsterite mantles (open markers).
  These results are all presented numerically in Table~\ref{tab:X_results}.
  \label{fig:scaling_law}}
\end{figure*}

\section{Results and Discussion} \label{sec:results}

The overall features of these giant impacts vary widely between scenarios, 
but continue to display the same range of behaviours and erosion mechanisms
that was examined in detail by \citetalias{Kegerreis+2020}.
Some of the possible outcomes are illustrated in Fig.~\ref{fig:demo_snaps},
with the particles that will become gravitationally unbound
in the rest frame of the target's core highlighted in purple.
The rows feature:
(1) a fast, head-on collision of our smallest impactor onto a small target,
resulting in near-total atmospheric loss and significant mantle erosion;
(2) a highly grazing impact leaving the target relatively undisturbed
while the impactor escapes;
(3) a slow, grazing impact of an equal-mass target and impactor,
significantly disrupting the planet but not violently enough 
to actually eject much unbound atmosphere;
and (4) a mid-angle collision onto a large target,
causing about half of the atmosphere to escape the system 
along with about half of the impactor.

\subsection{Erosion Trends} \label{sec:results:trends}

For a fixed impactor and target,
\citetalias{Kegerreis+2020} showed that the fraction of lost atmosphere 
scales as a simple function of the speed and impact parameter.
The most important missing pieces are the masses of the target and impactor.
We find that the atmospheric erosion depends neatly 
on the impactor:total mass ratio,
as shown in Fig.~\ref{fig:X_trends}~(left).
Furthermore, the fractional loss has no systematic dependence 
on the target (or total) mass
as long as the impactor mass ratio is the same.
These results continue to hold for
larger impactors hitting smaller targets 
and for bodies with different compositions and densities.

The slow, head-on scenarios (blue lines) show significant scatter.
This is consistent with the tests in \citetalias{Kegerreis+2020}
that showed how chaotic this specific type of collision can be,
unlike grazing or faster impacts.
Even tiny changes in the initial conditions 
can affect the details of the fall-back and sloshing that occurs 
after the initial impact and the resulting erosion.
This sets a relative uncertainty for these 
slow, head-on loss estimates of about 20\%.

Note that the mass ratio is not varied truly in isolation.
Although the other input parameters are kept constant,
the speed is set in terms of the escape speed and 
the angle in terms of the geometry of the system,
which depend on the body masses and radii 
and thus change along with the masses.

We find a similar dependence on the impact angle 
to that seen by \citetalias{Kegerreis+2020},
shown in Fig.~\ref{fig:X_trends}~(middle)
for the second suite of scenarios,
including the complex non-monotonic behaviour at low angles 
for slow and smaller impactors.
They found that a simple estimate of 
the fractional volume of the two bodies that interacts 
can account for any impact angle
across the full range of head-on to highly grazing collisions.
For the variable bulk densities in this study, 
we make the minor change to a fractional interacting mass, $f_M(b)$,
as detailed in Appx.~\ref{sec:appx:scaling},
though this modification makes little quantitative difference.
These results again appear to be completely independent 
of the total mass of the system.

The compositions, densities, 
and internal structures of the planets 
might also be expected affect the atmosphere loss.
With the third suite we test the extreme `terrestrial' cases 
of undifferentiated pure-iron and pure-rock bodies,
keeping either the mass or the radius the same as the standard versions.
The overall trends of this suite with the ratio of bulk densities 
(not including the atmosphere)
are shown in Fig.~\ref{fig:X_trends}~(right).
The mass ratios and the escape speeds also differ across these scenarios,
so it is unsurprising that no perfect scaling appears immediately.
Nonetheless, it is promising that 
some parameterisation of the density ratio 
could align the results across this highly diverse range of 
bodies and material combinations to a single trend,
once the mass and other parameters are accounted for.
Fig.~\ref{fig:X_trends}~(left) also confirms that 
these targets and impactors of very different compositions (yellow lines)
still follow the same neat scaling with the mass ratio as the standard cases.

\subsection{Scaling Law} \label{sec:results:scaling_law}

We find that the following power law
describes the fraction of eroded atmosphere 
from any impact scenario across this broad regime, 
as shown in Fig.~\ref{fig:scaling_law}:
\begin{equation}
  X \approx 0.64 \,\left[ \left(\dfrac{v_{\rm c}}{v_{\rm esc}}\right)^2 \,
    \left(\dfrac{M_{\rm i}}{M_{\rm tot}}\right)^{\tfrac{1}{2}} \,
    \left(\dfrac{\rho_{\rm i}}{\rho_{\rm t}}\right)^{\tfrac{1}{2}} \,
    f_M(b) \right]^{0.65} \;,
  \label{eqn:scaling_law}
\end{equation}
capped at 1 for total erosion.
The prefactor and exponent were found from a least-squares fit 
to the data points below total loss, with uncertainties of 0.01.

In spite of the mass and composition differences 
between the target and impactor planets
plus the dramatic qualitative differences between
slow, fast, head-on, grazing, and intermediate scenarios,
the median fractional deviation of the simulated loss fractions 
from the scaling law is only 9\%.
The ubiquitous independence of the loss on the total system mass 
is illustrated by the overlapping clusters of 
same-colour, different-shape points
around the scaling line in Fig.~\ref{fig:scaling_law}~(left).

We find that the specific impact energy 
$\left(\tfrac{1}{2} \mu v_{\rm c}^2 / M_{\rm tot}\right)$
is not the most convenient basis for a general scaling law.
Instead, normalising the speed at contact by the mutual escape speed
allows scenarios with different masses and densities
to be aligned by relatively simple additional terms.
The scenarios in \citetalias{Kegerreis+2020} also fit this trend well 
with a similar scatter to those in this study.

Scenarios with $b=0.7$ ($\beta = 44^\circ$)
show the tightest fit to the scaling law
(see also Fig.~\ref{fig:scaling_law_b_v}).
This is encouraging as $45^\circ$ is the most common angle for a collision.
The greatest discrepancies 
arise from some of the (less common) head-on or highly grazing impacts,
for which the loss changes with the scaling parameters 
more and less rapidly than the average trend, respectively.
The slow, head-on scenarios also suffer 
from their significant chaotic uncertainty.
To improve the fit for all angles would require that 
the power-law gradient be dependent on the angle.
However, this would yield only a minor improvement 
to the already reasonable fit at the cost of losing the current simplicity. 

The scaling law continues to agree well with the simulations that used 
the more sophisticated ANEOS equation of state (EoS),
as shown in Fig.~\ref{fig:scaling_law}~(right).
The median fractional difference in the atmospheric loss
from the equivalent scenarios simulated using the Tillotson EoS is 2\%.

Adding a thin atmosphere to the impactor 
does not affect significantly the fraction eroded from the target's atmosphere
(Fig.~\ref{fig:scaling_law},~right).
Furthermore, the scaling law still holds 
when the impactor is significantly more massive than the target.

The Moon-forming impact could have directly removed 
around 10 to 60\% of an atmosphere 
across a range of plausible scenarios. 
In order of increasing loss, 
the example impacts shown in Fig.~\ref{fig:scaling_law}~(left) are: 
canonical \citep{Canup+Asphaug2001},
hit-and-run \citep[][Fig.~1a]{Reufer+2012},
large impactor \citep[][Fig.~1]{Canup2012},
fast-spinning Earth \citep[][Fig.~1]{Cuk+Stewart2012},
and synestia \citep[][Fig.~7]{Lock+2018}.

\begin{figure}[t]
	\centering
	\includegraphics[width=0.9\columnwidth, trim={0mm 0mm 0mm 0mm}, clip]{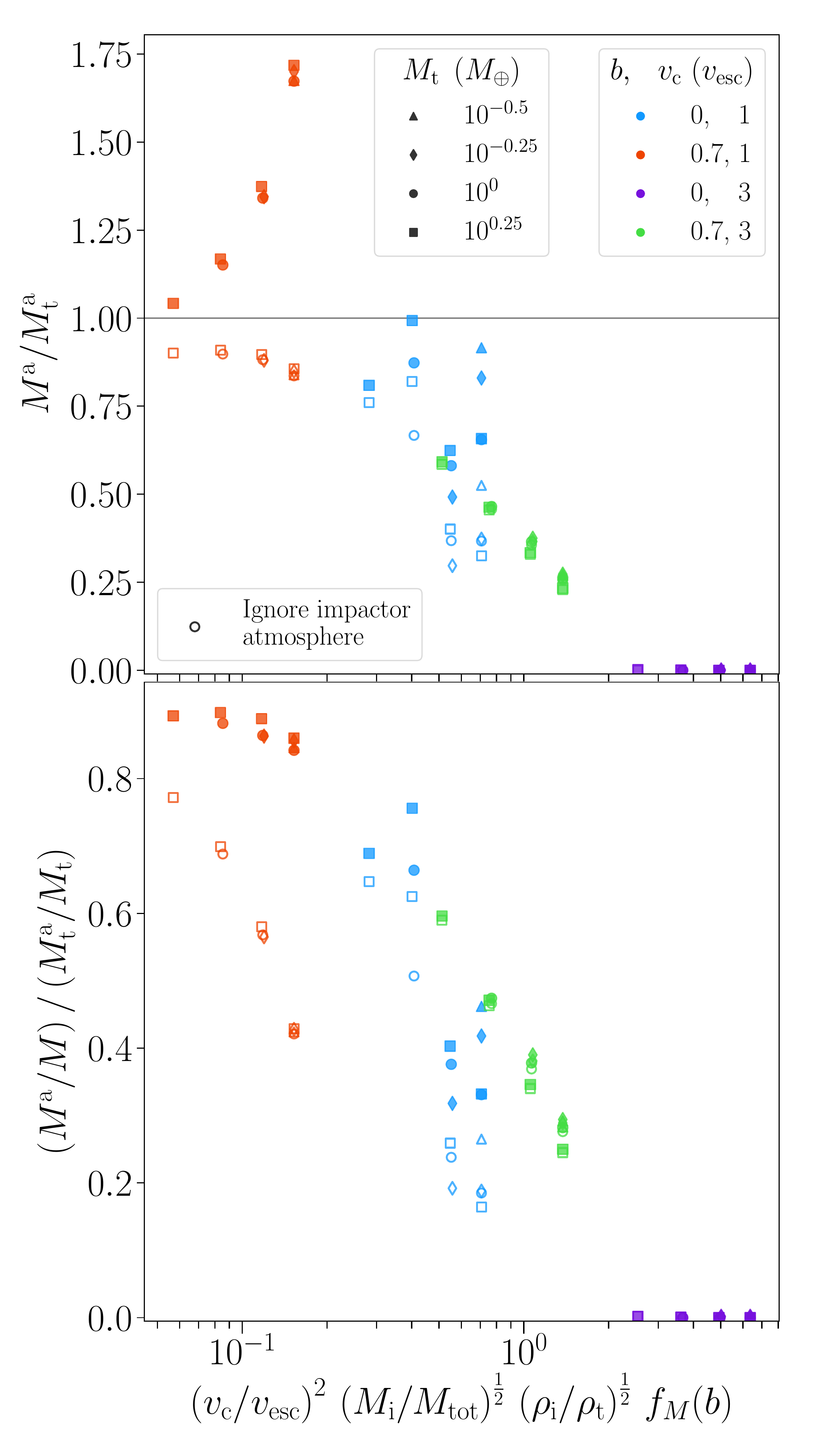}
  \caption{
    {\bf (top)} The final bound atmosphere mass relative to 
    the initial atmosphere mass of the target
    as a function of the scaling parameters,
    for scenarios with impactors that have 0.01$M_{\rm i}$ atmospheres,
    coloured by the impact parameter and speed
    with markers set by the target mass.
    Open markers ignore the 
    contribution of any atmosphere added by the impactor.
    {\bf (bottom)} The final bound atmosphere mass 
    as a fraction of the final core and mantle mass,
    relative to its initial value for the target.
		\label{fig:D_scaling}}
\end{figure}

\subsection{Volatile Delivery by Atmosphere-Hosting Impactors}

If the impactor also has an atmosphere, 
then some may survive delivery to the final planet.
For slow, grazing collisions the target can even end up with 
a larger atmosphere than it started with,
typically $\sim$85\% of the combined mass of both initial atmospheres
in these examples,
as shown in Fig.~\ref{fig:D_scaling}~(top).
Slow, head-on impacts are less generous, 
but a large proportion of the atmosphere's final composition
can still come from the impactor.
In the other subsets of much faster collisions tested here
either the grazing impactor escapes the system along with most of its atmosphere 
or the entirety of both atmospheres are ejected regardless.

This limited set of scenarios demonstrates that giant impacts can 
significantly build as well as erode an atmosphere, 
but further study is required to make robust predictions 
across a wider range of scenarios 
and different initial masses for both atmospheres.

However, the relative atmosphere mass always decreases
as a fraction of the planet's total mass,
as shown in Fig.~\ref{fig:D_scaling}~(bottom).
This demonstrates that although more atmosphere can be added than is removed,
even more mantle and/or core material is added in any scenario.
For the slow impactors that deliver significant atmosphere, 
$\sim$99\% of their core and mantle are also accreted.
Planets thus inevitably end up with a smaller mass fraction of atmosphere 
following this kind of impact.

\section{Conclusions: Applicability and Limitations} \label{sec:conc}

We have presented 3D simulations 
of giant impacts onto a range of terrestrial planets with thin atmospheres, 
including different masses, compositions and bulk densities,
equations of state, speeds, and impact angles.
We found a scaling law to estimate 
the fraction of atmosphere lost from any collision in this regime 
(Eqn.~\ref{eqn:scaling_law}).

This scaling law has been shown to hold empirically for 
target and impactor masses ranging from roughly three times the Earth's mass 
down to 0.3 and 0.05 of its mass, respectively;
for differentiated and undifferentiated planets
with densities from about half to over double the Earth's density;
and for any angle and speed.
The atmospheric erosion is independent of the system mass 
for a fixed ratio of the impactor and target masses.
Using the new ANEOS forsterite equation of state 
for the planets' mantles instead of the crude Tillotson 
has a negligible effect on the resulting loss.
We found no evident departure of the results from the trend 
at the extremes of these ranges, 
so it is plausible that the scaling applicability extends somewhat beyond.

The primary limitation for using this scaling law 
to make precise predictions elsewhere 
in the vast parameter space of giant impacts 
is the dependence on the atmosphere mass.
\citet[][\citetalias{Kegerreis+2020}]{Kegerreis+2020} 
found that the initial atmosphere mass has only a mild effect on the erosion
in this regime of `thin' atmospheres,
with $10\times$ lower mass leading to $\sim$10\% greater loss 
in their limited tests.
It is possible that this trend could be accounted for 
with an extra term in the scaling law,
but more focused study is required.
Thicker atmospheres that are able to significantly cushion the impactor 
and alter its trajectory might require a more different scaling approach.

The temperature of the atmosphere is also relevant, 
and \citetalias{Kegerreis+2020} found a similarly mild increase in loss
for 1500~K warmer atmospheres.
Comparable effects may be expected for different atmospheric compositions
to the H--He used here, which would similarly affect the scale height.
The longer-term thermal effects of a collision may cause additional loss
\citep{Biersteker+Schlichting2019},
the presence of an ocean beneath the atmosphere can increase the erosion
\citep{Genda+Abe2005},
and pre-impact rotation of the impactor and target 
could also cause significant differences \citep{Ruiz-Bonilla+2020}.

Giant impacts can readily remove anywhere from 
almost none to all of an atmosphere.
The strongest dependencies are on the angle and speed,
as well as the masses of both bodies and, to a lesser extent, their densities.
Slow impactors can also deliver a significant mass of atmosphere,
but always accompanied by larger proportions of their mantle and core.
Violent impacts can also erode the target's mantle,
typically removing at least $\sim$20\% for total atmospheric loss.
Different Moon-forming impact scenarios correspond to the direct loss of
from $\sim$10 to 60\% of a primordial atmosphere. 
This provides a new consideration for hypotheses of the Moon's origin 
in combination with models for the history of Earth's atmosphere.

Now that simulations like those presented here 
can be run with a high enough resolution to model 
the erosion of low-density atmospheres,
future studies can probe the remaining unexplored regimes 
and investigate the impacts of smaller and even larger bodies.
This way, robust scaling laws can continue to be built up 
to cover the full range of relevant scenarios 
in both our solar system and exoplanet systems
for the loss and delivery of volatiles by giant impacts.

\acknowledgments

We thank the anonymous reviewer for their highly constructive comments.
The research in this letter made use of the SWIFT open-source simulation code
\citep[\href{http://www.swiftsim.com}{www.swiftsim.com},][]{Schaller+2018}.
This work was supported by the Science and Technology Facilities Council (STFC)
grants ST/P000541/1 and ST/T000244/1, 
and used the DiRAC Data Centric system at Durham University,
operated by the Institute for Computational Cosmology on behalf of the
STFC DiRAC HPC Facility (www.dirac.ac.uk).
This equipment was funded by
BIS National E-infrastructure capital grant ST/K00042X/1,
STFC capital grants ST/H008519/1 and ST/K00087X/1,
STFC DiRAC Operations grant ST/K003267/1 and Durham University.
DiRAC is part of the National E-Infrastructure.
JAK acknowledges support from STFC grants ST/N001494/1 and ST/T002565/1.
DCC, LFTA, and KJZ acknowledge support from 
NASA Planetary Atmospheres grant NNX14AJ45G.
RJM is supported by the Royal Society.

%



\software{
  SWIFT (\href{www.swiftsim.com}{www.swiftsim.com},
  \citet{Kegerreis+2019}, \citet{Schaller+2016}, version 0.8.5);
  WoMa (\href{https://pypi.org/project/woma/}{pypi.org/project/woma/},
  \citet{Ruiz-Bonilla+2020}).
}



\clearpage
\appendix
\renewcommand\thefigure{\thesection\arabic{figure}}

\setcounter{figure}{0}

\section{Initial Conditions and Impact Scenarios} \label{sec:appx:methods}

\begin{figure}[t]
	\centering
	\includegraphics[width=0.9\columnwidth, trim={8mm 9mm 11mm 6mm}, clip]{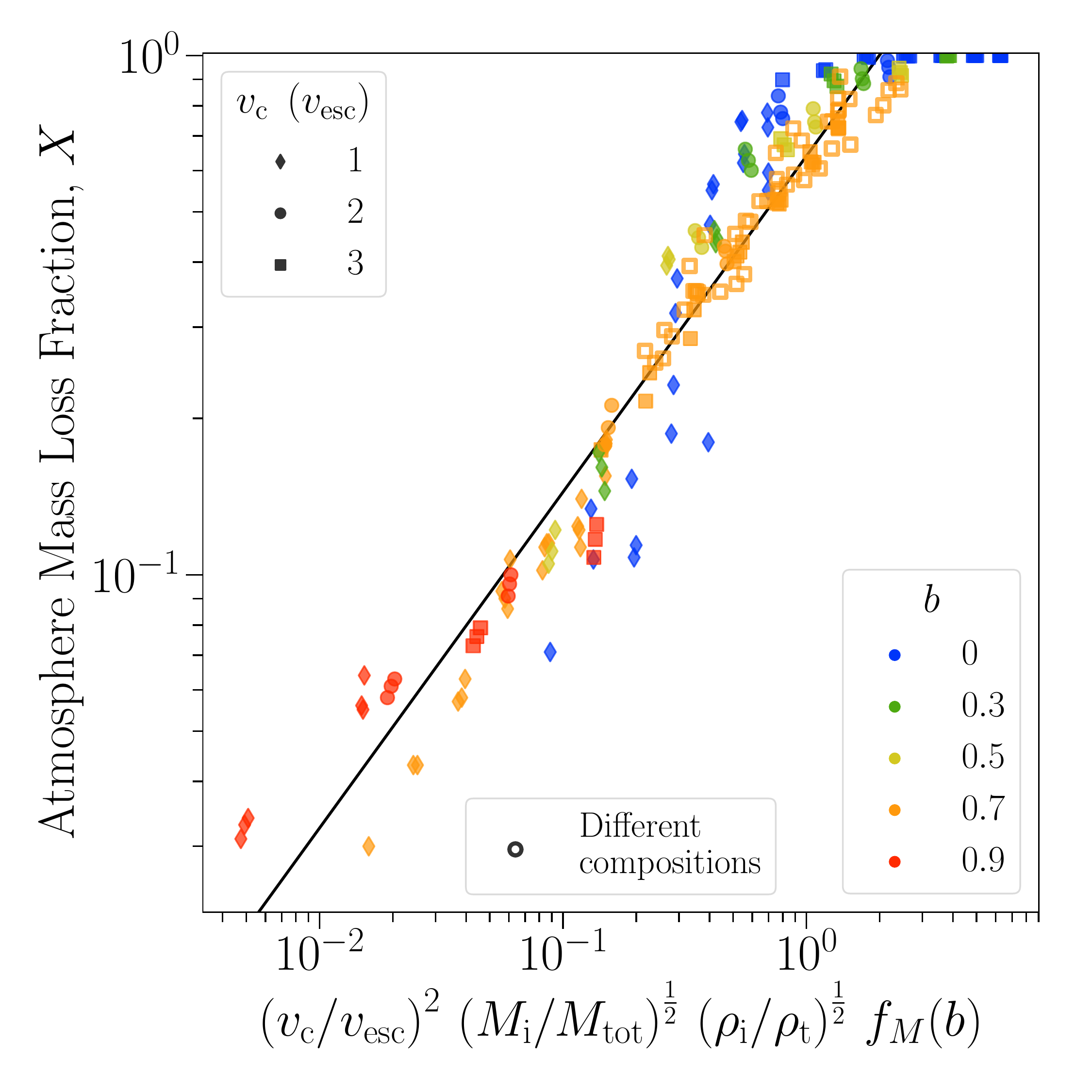}
  \caption{
    Same as Fig.~\ref{fig:scaling_law}~(left) 
    but coloured instead by the impact parameter
    with markers set by the speed at contact.
		\label{fig:scaling_law_b_v}}
\end{figure}

The input parameters for each simulation are listed in 
Tables~\ref{tab:masses_and_radii}, \ref{tab:X_results}, \ref{tab:X_results_2}, 
and \ref{tab:X_results_3},
along with the resulting atmospheric erosion shown in 
Figs.~\ref{fig:scaling_law} and \ref{fig:scaling_law_b_v}.
Table~\ref{tab:D_results} lists the data shown in Fig.~\ref{fig:D_scaling}.

For the first suite of changing masses, 
our four targets have masses of
$10^{-0.5,\,-0.25,\,0,\,0.25}$~$M_\oplus$,
not including the atmospheres,
with up to seven impactor masses between 
$10^{-1.25}$~$M_\oplus$ and the target's mass
with the same logarithmic spacing of $0.25$ dex,
for a total of 22 target and impactor combinations.
Table~\ref{tab:masses_and_radii} lists these masses and corresponding radii.
Each combination is simulated in
four scenarios: head-on, grazing, slow, and fast
-- $b=0,0.7$ and $v_c=1,3~v_{\rm esc}$ --
for a total of 88 simulations.
For the four impactors with mass $\geq 10^{-0.5}~M_\oplus$
we also run a duplicate simulation where the impactor also has an 
added atmosphere of 1\% of its mass,
for an extra 40 simulations.
Furthermore, these atmosphere-hosting impactors 
can also be treated as the targets.
This provides an additional set of scenarios for 
erosion by impactors that are more massive than the target.

\begin{table}[b]
	\begin{center} \begin{tabular}{lllllll}
    \cline{0-6}
    \multicolumn{3}{c}{Standard} & \multicolumn{2}{c}{Same-Mass} 
    & \multicolumn{2}{c}{Same-Radius} \\
    \multicolumn{2}{c}{Mass} & Radius & Iron & Rock & Iron & Rock \\
    $(M_\oplus)$ & $(M_\oplus)$ & $(R_\oplus)$ & $(R_\oplus)$ 
    & $(R_\oplus)$ & $(M_\oplus)$ & $(M_\oplus)$ \\
    \cline{0-6}
    $10^{-1.25}$ & 0.056 & 0.444 \\
    $10^{-1}$ & 0.100 & 0.538 & 0.397 & 0.568 & 0.260 & 0.0782 \\
    $10^{-0.75}$ & 0.178 & 0.625 \\
    $10^{-0.5}$ & 0.316 & 0.733 & 0.559 & 0.788 & 0.844 & 0.245 \\
    $10^{-0.25}$ & 0.562 & 0.856 \\
    $10^{0}$ & 1.000 & 0.992 & 0.768 & 1.062 & 2.715 & 0.766 \\
    $10^{0.25}$ & 1.778 & 1.153 \\
    \cline{0-6}
	\end{tabular} \end{center}
	\caption{The masses and radii of the bodies, ignoring any atmosphere.
    The bodies for the different-density suite are composed of pure iron or rock 
    and either the standard mass or radius is kept the same,
    giving a new radius or mass, respectively.
    \label{tab:masses_and_radii}}
\end{table}

For the second suite of changing speeds and angles, 
we select the impactors that are less massive than each target 
by 1 and 0.25 dex (with no atmospheres)
for the three larger targets.
In other words, the following six mass combinations (in $M_\oplus$) are used:
$10^{-0.25}$ and $10^{-1.25,\,-0.5}$;
$10^{0}$ and $10^{-1,\,-0.25}$;
$10^{0.25}$ and $10^{-0.75,\,0}$.
Each combination is simulated in scenarios with 
impact parameter $b = 0, 0.3, 0.5, 0.7, 0.9$ 
and speed at contact $v_{\rm c} = 1, 2, 3$~$v_{\rm esc}$
for a total of 90 simulations,
out of which 24 are duplicates of the first suite.

For the third suite of different-density bodies,
we take as a base a fast, grazing scenario 
($b=0.7$, $v_{\rm c} = 3$~$v_{\rm esc}$)
with the $10^0$~$M_\oplus$ target and $10^{-1, -0.5, 0}$~$M_\oplus$ impactors.
These collisions yield middling erosion 
and tend to align closely with previous scaling laws \citep{Kegerreis+2020}.
For each of these default planets, we create new versions
that are made entirely of iron or entirely of rock
(instead of the default 30:70 mass ratio)
keeping the same masses and allowing the radii to change,
or keeping the same radii and allowing the masses to change,
as listed in Table~\ref{tab:masses_and_radii}.
We simulate the collision of each impactor with each target
(skipping some combinations for the smallest and largest impactor, 
as detailed in Table~\ref{tab:X_results_3}),
for a total of 47 simulations,
out of which three are duplicates from the first two suites.

Finally, we run 21 additional simulations using 
the new ANEOS forsterite \citep{Stewart+2019} instead of Tillotson
as the mantle material in both the targets and impactors.
We collide $10^{-1.25, -0.75, -0.25}$~$M_\oplus$ impactors
with the $10^0$~$M_\oplus$ target,
for $b=0.7$ with $v_{\rm c} = 1, 2, 3$~$v_{\rm esc}$,
and $b=0, 0.3, 0.5, 0.7, 0.9$ with $v_{\rm c} = 2$~$v_{\rm esc}$.

To set the number of SPH particles in each simulation,
for the smaller two targets 
we use $10^7$ particles per $10^{-0.5}$~$M_\oplus$
and for the larger two we use $10^7$ particles per $M_\oplus$,
giving particle masses of $3.2 \times 10^{-8}~M_\oplus = 1.9 \times 10^{17}$~kg
and $10^{-7}~M_\oplus = 6.0 \times 10^{17}$~kg, respectively.
This avoids the otherwise insufficient or unnecessarily high resolution
for the smallest and largest targets
if we had instead chosen a single particle mass throughout.
The small downside is that two versions of most impactors must be created 
to match the particle mass of the target in each case.

In order to run the simulations 
until the amount of eroded material 
no longer changes significantly \citep[see][Fig.~6]{Kegerreis+2020},
high-speed and/or low-angle scenarios with $v_{\rm c} = 3$,
or $v_{\rm c} = 2$ and $b = 0, 0.3$,
are run for 5~hours after contact, 
the others are run conservatively for 14~hours
(plus the initial 1~hour before contact in both cases).
The three simulations with $b=0.9$, $v_{\rm c}=1$~$v_{\rm esc}$, 
and an impactor:target mass ratio $10^{-0.25}$
are exceptions and are stopped (in terms of their analysis) after 8.5~h,
before the nearly-intact impactor fragment re-collides with the target.
The double impacts in these unusual cases must be treated as separate collisions
in order to follow the scaling law as any other scenario.
Snapshots of the particle data are output every 500~s.


Most simulations are run in the centre-of-mass and zero-momentum frame.
The exceptions are the high-speed, grazing impacts with massive impactors.
The targets in these scenarios would rapidly exit the simulation box 
(of side length 80~$R_\oplus$)
as the unbound impactors fly out the opposite side.
To avoid this, the following small subset of simulations are 
run instead in the initial rest frame of the target
and in a larger 120~$R_\oplus$ box:
if (1) the impactor is either the same mass as the target or
-- for the larger two targets -- 0.25 dex less massive;
and (2) $v_{\rm c} \ge 2$ with $b \ge 0.5$.

\begin{table*}[t]
  \small
	\begin{center} \begin{tabular}{lllcc r lllcc r lllcc}
    \cline{0-4} \cline{7-11} \cline{13-17}
    $M_{\rm t}$ & $M_{\rm i}$ & $b$ & $v_{\rm c}$ & $X$ & 
    & $M_{\rm t}$ & $M_{\rm i}$ & $b$ & $v_{\rm c}$ & $X$ & 
    & $M_{\rm t}$ & $M_{\rm i}$ & $b$ & $v_{\rm c}$ & $X$ \\
    $\left(M_\oplus\right)$ & $\left(M_\oplus\right)$ & & $\left(v_{\rm esc}\right)$ & &
    & $\left(M_\oplus\right)$ & $\left(M_\oplus\right)$ & & $\left(v_{\rm esc}\right)$ & &
    & $\left(M_\oplus\right)$ & $\left(M_\oplus\right)$ & & $\left(v_{\rm esc}\right)$ & \\
		\cline{0-4} \cline{7-11} \cline{13-17}
    \multicolumn{5}{c}{\emph{first suite}} & 
    & $10^{-0.25}$ & $10^{-0.25}$ & 0.7 & 3 & 0.723 & 
    & $10^{0.25}$ & $10^{-0.75}$ & 0.7 & 3 & 0.285 \\ 
    $10^{-0.5}$ & $10^{-1.25}$ & 0 & 1 & 0.372 & 
    & $10^{0}$ & $10^{-1.25}$ & 0 & 1 & 0.107 & 
    & $10^{0.25}$ & $10^{-0.5}$ & 0 & 1 & 0.187 \\ 
    $10^{-0.5}$ & $10^{-1.25}$ & 0.7 & 1 & 0.107 & 
    & $10^{0}$ & $10^{-1.25}$ & 0.7 & 1 & 0.043 & 
    & $10^{0.25}$ & $10^{-0.5}$ & 0.7 & 1 & 0.093 \\ 
    $10^{-0.5}$ & $10^{-1.25}$ & 0 & 3 & 0.997 & 
    & $10^{0}$ & $10^{-1.25}$ & 0 & 3 & 0.939 & 
    & $10^{0.25}$ & $10^{-0.5}$ & 0 & 3 & 0.998 \\ 
    $10^{-0.5}$ & $10^{-1.25}$ & 0.7 & 3 & 0.437 & 
    & $10^{0}$ & $10^{-1.25}$ & 0.7 & 3 & 0.245 & 
    & $10^{0.25}$ & $10^{-0.5}$ & 0.7 & 3 & 0.401 \\ 
    $10^{-0.5}$ & $10^{-1}$ & 0 & 1 & 0.565 & 
    & $10^{0}$ & $10^{-1}$ & 0 & 1 & 0.108 & 
    & $10^{0.25}$ & $10^{-0.25}$ & 0 & 1 & 0.180 \\ 
    $10^{-0.5}$ & $10^{-1}$ & 0.7 & 1 & 0.115 & 
    & $10^{0}$ & $10^{-1}$ & 0.7 & 1 & 0.058 & 
    & $10^{0.25}$ & $10^{-0.25}$ & 0.7 & 1 & 0.102 \\ 
    $10^{-0.5}$ & $10^{-1}$ & 0 & 3 & 1.000 & 
    & $10^{0}$ & $10^{-1}$ & 0 & 3 & 0.997 & 
    & $10^{0.25}$ & $10^{-0.25}$ & 0 & 3 & 1.000 \\ 
    $10^{-0.5}$ & $10^{-1}$ & 0.7 & 3 & 0.527 & 
    & $10^{0}$ & $10^{-1}$ & 0.7 & 3 & 0.324 & 
    & $10^{0.25}$ & $10^{-0.25}$ & 0.7 & 3 & 0.528 \\ 
    $10^{-0.5}$ & $10^{-0.75}$ & 0 & 1 & 0.646 & 
    & $10^{0}$ & $10^{-0.75}$ & 0 & 1 & 0.232 & 
    & $10^{0.25}$ & $10^{0}$ & 0 & 1 & 0.745 \\ 
    $10^{-0.5}$ & $10^{-0.75}$ & 0.7 & 1 & 0.140 & 
    & $10^{0}$ & $10^{-0.75}$ & 0.7 & 1 & 0.090 & 
    & $10^{0.25}$ & $10^{0}$ & 0.7 & 1 & 0.124 \\ 
    $10^{-0.5}$ & $10^{-0.75}$ & 0 & 3 & 1.000 & 
    & $10^{0}$ & $10^{-0.75}$ & 0 & 3 & 0.998 & 
    & $10^{0.25}$ & $10^{0}$ & 0 & 3 & 1.000 \\ 
    $10^{-0.5}$ & $10^{-0.75}$ & 0.7 & 3 & 0.623 & 
    & $10^{0}$ & $10^{-0.75}$ & 0.7 & 3 & 0.411 & 
    & $10^{0.25}$ & $10^{0}$ & 0.7 & 3 & 0.653 \\ 
    $10^{-0.5}$ & $10^{-0.5}$ & 0 & 1 & 0.595 & 
    & $10^{0}$ & $10^{-0.5}$ & 0 & 1 & 0.472 & 
    & $10^{0.25}$ & $10^{0.25}$ & 0 & 1 & 0.776 \\ 
    $10^{-0.5}$ & $10^{-0.5}$ & 0.7 & 1 & 0.182 & 
    & $10^{0}$ & $10^{-0.5}$ & 0.7 & 1 & 0.113 & 
    & $10^{0.25}$ & $10^{0.25}$ & 0.7 & 1 & 0.155 \\ 
    $10^{-0.5}$ & $10^{-0.5}$ & 0 & 3 & 1.000 & 
    & $10^{0}$ & $10^{-0.5}$ & 0 & 3 & 1.000 & 
    & $10^{0.25}$ & $10^{0.25}$ & 0 & 3 & 1.000 \\ 
    $10^{-0.5}$ & $10^{-0.5}$ & 0.7 & 3 & 0.726 & 
    & $10^{0}$ & $10^{-0.5}$ & 0.7 & 3 & 0.520 & 
    & $10^{0.25}$ & $10^{0.25}$ & 0.7 & 3 & 0.758 \\ 
    \cdashline{13-17}
    $10^{-0.25}$ & $10^{-1.25}$ & 0 & 1 & 0.114 & 
    & $10^{0}$ & $10^{-0.25}$ & 0 & 1 & 0.751 & 
    & \multicolumn{5}{c}{\emph{second suite}} \\ 
    $10^{-0.25}$ & $10^{-1.25}$ & 0.7 & 1 & 0.063 & 
    & $10^{0}$ & $10^{-0.25}$ & 0.7 & 1 & 0.122 & 
    & $10^{-0.25}$ & $10^{-1.25}$ & 0.3 & 1 & 0.145 \\ 
    $10^{-0.25}$ & $10^{-1.25}$ & 0 & 3 & 0.992 & 
    & $10^{0}$ & $10^{-0.25}$ & 0 & 3 & 1.000 & 
    & $10^{-0.25}$ & $10^{-1.25}$ & 0.5 & 1 & 0.122 \\ 
    $10^{-0.25}$ & $10^{-1.25}$ & 0.7 & 3 & 0.347 & 
    & $10^{0}$ & $10^{-0.25}$ & 0.7 & 3 & 0.624 & 
    & $10^{-0.25}$ & $10^{-1.25}$ & 0.9 & 1 & 0.034 \\ 
    $10^{-0.25}$ & $10^{-1}$ & 0 & 1 & 0.319 & 
    & $10^{0}$ & $10^{0}$ & 0 & 1 & 0.727 & 
    & $10^{-0.25}$ & $10^{-1.25}$ & 0 & 2 & 0.756 \\ 
    $10^{-0.25}$ & $10^{-1}$ & 0.7 & 1 & 0.086 & 
    & $10^{0}$ & $10^{0}$ & 0.7 & 1 & 0.177 & 
    & $10^{-0.25}$ & $10^{-1.25}$ & 0.3 & 2 & 0.601 \\ 
    $10^{-0.25}$ & $10^{-1}$ & 0 & 3 & 0.997 & 
    & $10^{0}$ & $10^{0}$ & 0 & 3 & 1.000 & 
    & $10^{-0.25}$ & $10^{-1.25}$ & 0.5 & 2 & 0.427 \\ 
    $10^{-0.25}$ & $10^{-1}$ & 0.7 & 3 & 0.418 & 
    & $10^{0}$ & $10^{0}$ & 0.7 & 3 & 0.726 & 
    & $10^{-0.25}$ & $10^{-1.25}$ & 0.7 & 2 & 0.212 \\ 
    $10^{-0.25}$ & $10^{-0.75}$ & 0 & 1 & 0.550 & 
    & $10^{0.25}$ & $10^{-1.25}$ & 0 & 1 & 0.071 & 
    & $10^{-0.25}$ & $10^{-1.25}$ & 0.9 & 2 & 0.063 \\ 
    $10^{-0.25}$ & $10^{-0.75}$ & 0.7 & 1 & 0.115 & 
    & $10^{0.25}$ & $10^{-1.25}$ & 0.7 & 1 & 0.030 & 
    & $10^{-0.25}$ & $10^{-1.25}$ & 0.3 & 3 & 0.872 \\ 
    $10^{-0.25}$ & $10^{-0.75}$ & 0 & 3 & 1.000 & 
    & $10^{0.25}$ & $10^{-1.25}$ & 0 & 3 & 0.898 & 
    & $10^{-0.25}$ & $10^{-1.25}$ & 0.5 & 3 & 0.658 \\ 
    $10^{-0.25}$ & $10^{-0.75}$ & 0.7 & 3 & 0.518 & 
    & $10^{0.25}$ & $10^{-1.25}$ & 0.7 & 3 & 0.174 & 
    & $10^{-0.25}$ & $10^{-1.25}$ & 0.9 & 3 & 0.079 \\ 
    $10^{-0.25}$ & $10^{-0.5}$ & 0 & 1 & 0.621 & 
    & $10^{0.25}$ & $10^{-1}$ & 0 & 1 & 0.134 & 
    & $10^{-0.25}$ & $10^{-0.5}$ & 0.3 & 1 & 0.444 \\ 
    $10^{-0.25}$ & $10^{-0.5}$ & 0.7 & 1 & 0.113 & 
    & $10^{0.25}$ & $10^{-1}$ & 0.7 & 1 & 0.043 & 
    & $10^{-0.25}$ & $10^{-0.5}$ & 0.5 & 1 & 0.405 \\ 
    $10^{-0.25}$ & $10^{-0.5}$ & 0 & 3 & 1.000 & 
    & $10^{0.25}$ & $10^{-1}$ & 0 & 3 & 0.935 & 
    & $10^{-0.25}$ & $10^{-0.5}$ & 0.9 & 1 & 0.064 \\ 
    $10^{-0.25}$ & $10^{-0.5}$ & 0.7 & 3 & 0.616 & 
    & $10^{0.25}$ & $10^{-1}$ & 0.7 & 3 & 0.216 & 
    & $10^{-0.25}$ & $10^{-0.5}$ & 0 & 2 & 0.910 \\ 
    $10^{-0.25}$ & $10^{-0.25}$ & 0 & 1 & 0.550 & 
    & $10^{0.25}$ & $10^{-0.75}$ & 0 & 1 & 0.153 & 
    & $10^{-0.25}$ & $10^{-0.5}$ & 0.3 & 2 & 0.883 \\ 
    $10^{-0.25}$ & $10^{-0.25}$ & 0.7 & 1 & 0.179 & 
    & $10^{0.25}$ & $10^{-0.75}$ & 0.7 & 1 & 0.057 & 
    & $10^{-0.25}$ & $10^{-0.5}$ & 0.5 & 2 & 0.728 \\ 
    $10^{-0.25}$ & $10^{-0.25}$ & 0 & 3 & 1.000 & 
    & $10^{0.25}$ & $10^{-0.75}$ & 0 & 3 & 0.997 & 
    & $10^{-0.25}$ & $10^{-0.5}$ & 0.7 & 2 & 0.397 \\ 
		\cline{0-4} \cline{7-11} \cline{13-17}
	\end{tabular} \end{center}
	\caption{The target mass, $M_{\rm t}$, impactor mass, $M_{\rm i}$, 
    impact parameter, $b$, speed at contact, $v_{\rm c}$,
    and lost mass fraction of the atmosphere, $X$, 
    for the simulation scenarios, 
    as presented in Fig.~\ref{fig:scaling_law}.
    The dashed line after the first 88 simulations 
    indicates the start of the second suite
    as described in Appx.~\ref{sec:appx:methods},
    not including any duplicates.
    Continued in Tables~\ref{tab:X_results_2} and \ref{tab:X_results_3}.
    These tables are available in machine-readable form from the online version.
    \label{tab:X_results}}
  \vspace{-1em}
\end{table*}
\begin{table*}[t]
  \small
	\begin{center} \begin{tabular}{lllcc r lllcc r lllcc}
    \cline{0-4} \cline{7-11} \cline{13-17}
    $M_{\rm t}$ & $M_{\rm i}$ & $b$ & $v_{\rm c}$ & $X$ & 
    & $M_{\rm t}$ & $M_{\rm i}$ & $b$ & $v_{\rm c}$ & $X$ & 
    & $M_{\rm t}$ & $M_{\rm i}$ & $b$ & $v_{\rm c}$ & $X$ \\
    $\left(M_\oplus\right)$ & $\left(M_\oplus\right)$ & & $\left(v_{\rm esc}\right)$ & &
    & $\left(M_\oplus\right)$ & $\left(M_\oplus\right)$ & & $\left(v_{\rm esc}\right)$ & &
    & $\left(M_\oplus\right)$ & $\left(M_\oplus\right)$ & & $\left(v_{\rm esc}\right)$ & \\
		\cline{0-4} \cline{7-11} \cline{13-17}
    $10^{-0.25}$ & $10^{-0.5}$ & 0.9 & 2 & 0.100 & 
    & $10^{0.25}$ & $10^{-0.75}$ & 0.9 & 3 & 0.073 & 
    & $10^{-0.5}$ & $^\star$$10^{-0.5}$ & 0.7 & 1 & 0.161 \\ 
    $10^{-0.25}$ & $10^{-0.5}$ & 0.3 & 3 & 0.999 & 
    & $10^{0.25}$ & $10^{0}$ & 0.3 & 1 & 0.461 & 
    & $10^{-0.5}$ & $^\star$$10^{-0.5}$ & 0 & 3 & 1.000 \\ 
    $10^{-0.25}$ & $10^{-0.5}$ & 0.5 & 3 & 0.915 & 
    & $10^{0.25}$ & $10^{0}$ & 0.5 & 1 & 0.394 & 
    & $10^{-0.5}$ & $^\star$$10^{-0.5}$ & 0.7 & 3 & 0.738 \\ 
    $10^{-0.25}$ & $10^{-0.5}$ & 0.9 & 3 & 0.125 & 
    & $10^{0.25}$ & $10^{0}$ & 0.9 & 1 & 0.056 & 
    & $10^{-0.25}$ & $^\star$$10^{-0.5}$ & 0 & 1 & 0.703 \\ 
    $10^{0}$ & $10^{-1}$ & 0.3 & 1 & 0.161 & 
    & $10^{0.25}$ & $10^{0}$ & 0 & 2 & 0.978 & 
    & $10^{-0.25}$ & $^\star$$10^{-0.5}$ & 0.7 & 1 & 0.120 \\ 
    $10^{0}$ & $10^{-1}$ & 0.5 & 1 & 0.111 & 
    & $10^{0.25}$ & $10^{0}$ & 0.3 & 2 & 0.943 & 
    & $10^{-0.25}$ & $^\star$$10^{-0.5}$ & 0 & 3 & 1.000 \\ 
    $10^{0}$ & $10^{-1}$ & 0.9 & 1 & 0.033 & 
    & $10^{0.25}$ & $10^{0}$ & 0.5 & 2 & 0.790 & 
    & $10^{-0.25}$ & $^\star$$10^{-0.5}$ & 0.7 & 3 & 0.634 \\ 
    $10^{0}$ & $10^{-1}$ & 0 & 2 & 0.778 & 
    & $10^{0.25}$ & $10^{0}$ & 0.7 & 2 & 0.429 & 
    & $10^{-0.25}$ & $^\star$$10^{-0.25}$ & 0 & 1 & 0.627 \\ 
    $10^{0}$ & $10^{-1}$ & 0.3 & 2 & 0.628 & 
    & $10^{0.25}$ & $10^{0}$ & 0.9 & 2 & 0.091 & 
    & $10^{-0.25}$ & $^\star$$10^{-0.25}$ & 0.7 & 1 & 0.151 \\ 
    $10^{0}$ & $10^{-1}$ & 0.5 & 2 & 0.446 & 
    & $10^{0.25}$ & $10^{0}$ & 0.3 & 3 & 0.998 & 
    & $10^{-0.25}$ & $^\star$$10^{-0.25}$ & 0 & 3 & 1.000 \\ 
    $10^{0}$ & $10^{-1}$ & 0.7 & 2 & 0.192 & 
    & $10^{0.25}$ & $10^{0}$ & 0.5 & 3 & 0.946 & 
    & $10^{-0.25}$ & $^\star$$10^{-0.25}$ & 0.7 & 3 & 0.732 \\ 
    $10^{0}$ & $10^{-1}$ & 0.9 & 2 & 0.061 & 
    & $10^{0.25}$ & $10^{0}$ & 0.9 & 3 & 0.108 & 
    & $10^{0}$ & $^\star$$10^{-0.5}$ & 0 & 1 & 0.333 \\ 
    \cdashline{7-11}
    $10^{0}$ & $10^{-1}$ & 0.3 & 3 & 0.894 & 
    & \multicolumn{5}{c}{\emph{ANEOS forsterite mantles}} & 
    & $10^{0}$ & $^\star$$10^{-0.5}$ & 0.7 & 1 & 0.102 \\ 
    $10^{0}$ & $10^{-1}$ & 0.5 & 3 & 0.673 & 
    & $^\dagger$$10^{0}$ & $^\dagger$$10^{-1.25}$ & 0.7 & 1 & 0.041 & 
    & $10^{0}$ & $^\star$$10^{-0.5}$ & 0 & 3 & 1.000 \\ 
    $10^{0}$ & $10^{-1}$ & 0.9 & 3 & 0.076 & 
    & $^\dagger$$10^{0}$ & $^\dagger$$10^{-1.25}$ & 0 & 2 & 0.513 & 
    & $10^{0}$ & $^\star$$10^{-0.5}$ & 0.7 & 3 & 0.543 \\ 
    $10^{0}$ & $10^{-0.25}$ & 0.3 & 1 & 0.435 & 
    & $^\dagger$$10^{0}$ & $^\dagger$$10^{-1.25}$ & 0.3 & 2 & 0.456 & 
    & $10^{0}$ & $^\star$$10^{-0.25}$ & 0 & 1 & 0.632 \\ 
    $10^{0}$ & $10^{-0.25}$ & 0.5 & 1 & 0.411 & 
    & $^\dagger$$10^{0}$ & $^\dagger$$10^{-1.25}$ & 0.5 & 2 & 0.349 & 
    & $10^{0}$ & $^\star$$10^{-0.25}$ & 0.7 & 1 & 0.118 \\ 
    $10^{0}$ & $10^{-0.25}$ & 0.9 & 1 & 0.055 & 
    & $^\dagger$$10^{0}$ & $^\dagger$$10^{-1.25}$ & 0.7 & 2 & 0.170 & 
    & $10^{0}$ & $^\star$$10^{-0.25}$ & 0 & 3 & 1.000 \\ 
    $10^{0}$ & $10^{-0.25}$ & 0 & 2 & 0.949 & 
    & $^\dagger$$10^{0}$ & $^\dagger$$10^{-1.25}$ & 0.9 & 2 & 0.056 & 
    & $10^{0}$ & $^\star$$10^{-0.25}$ & 0.7 & 3 & 0.644 \\ 
    $10^{0}$ & $10^{-0.25}$ & 0.3 & 2 & 0.902 & 
    & $^\dagger$$10^{0}$ & $^\dagger$$10^{-1.25}$ & 0.7 & 3 & 0.278 & 
    & $10^{0}$ & $^\star$$10^{0}$ & 0 & 1 & 0.633 \\ 
    $10^{0}$ & $10^{-0.25}$ & 0.5 & 2 & 0.745 & 
    & $^\dagger$$10^{0}$ & $^\dagger$$10^{-0.75}$ & 0.7 & 1 & 0.107 & 
    & $10^{0}$ & $^\star$$10^{0}$ & 0.7 & 1 & 0.164 \\ 
    $10^{0}$ & $10^{-0.25}$ & 0.7 & 2 & 0.420 & 
    & $^\dagger$$10^{0}$ & $^\dagger$$10^{-0.75}$ & 0 & 2 & 0.852 & 
    & $10^{0}$ & $^\star$$10^{0}$ & 0 & 3 & 1.000 \\ 
    $10^{0}$ & $10^{-0.25}$ & 0.9 & 2 & 0.096 & 
    & $^\dagger$$10^{0}$ & $^\dagger$$10^{-0.75}$ & 0.3 & 2 & 0.720 & 
    & $10^{0}$ & $^\star$$10^{0}$ & 0.7 & 3 & 0.743 \\ 
    $10^{0}$ & $10^{-0.25}$ & 0.3 & 3 & 0.999 & 
    & $^\dagger$$10^{0}$ & $^\dagger$$10^{-0.75}$ & 0.5 & 2 & 0.554 & 
    & $10^{0.25}$ & $^\star$$10^{-0.5}$ & 0 & 1 & 0.240 \\ 
    $10^{0}$ & $10^{-0.25}$ & 0.5 & 3 & 0.927 & 
    & $^\dagger$$10^{0}$ & $^\dagger$$10^{-0.75}$ & 0.7 & 2 & 0.254 & 
    & $10^{0.25}$ & $^\star$$10^{-0.5}$ & 0.7 & 1 & 0.099 \\ 
    $10^{0}$ & $10^{-0.25}$ & 0.9 & 3 & 0.117 & 
    & $^\dagger$$10^{0}$ & $^\dagger$$10^{-0.75}$ & 0.9 & 2 & 0.065 & 
    & $10^{0.25}$ & $^\star$$10^{-0.5}$ & 0 & 3 & 0.999 \\ 
    $10^{0.25}$ & $10^{-0.75}$ & 0.3 & 1 & 0.172 & 
    & $^\dagger$$10^{0}$ & $^\dagger$$10^{-0.75}$ & 0.7 & 3 & 0.402 & 
    & $10^{0.25}$ & $^\star$$10^{-0.5}$ & 0.7 & 3 & 0.415 \\ 
    $10^{0.25}$ & $10^{-0.75}$ & 0.5 & 1 & 0.105 & 
    & $^\dagger$$10^{0}$ & $^\dagger$$10^{-0.25}$ & 0.7 & 1 & 0.157 & 
    & $10^{0.25}$ & $^\star$$10^{-0.25}$ & 0 & 1 & 0.180 \\ 
    $10^{0.25}$ & $10^{-0.75}$ & 0.9 & 1 & 0.031 & 
    & $^\dagger$$10^{0}$ & $^\dagger$$10^{-0.25}$ & 0 & 2 & 0.926 & 
    & $10^{0.25}$ & $^\star$$10^{-0.25}$ & 0.7 & 1 & 0.091 \\ 
    $10^{0.25}$ & $10^{-0.75}$ & 0 & 2 & 0.836 & 
    & $^\dagger$$10^{0}$ & $^\dagger$$10^{-0.25}$ & 0.3 & 2 & 0.893 & 
    & $10^{0.25}$ & $^\star$$10^{-0.25}$ & 0 & 3 & 1.000 \\ 
    $10^{0.25}$ & $10^{-0.75}$ & 0.3 & 2 & 0.660 & 
    & $^\dagger$$10^{0}$ & $^\dagger$$10^{-0.25}$ & 0.5 & 2 & 0.749 & 
    & $10^{0.25}$ & $^\star$$10^{-0.25}$ & 0.7 & 3 & 0.545 \\ 
    $10^{0.25}$ & $10^{-0.75}$ & 0.5 & 2 & 0.460 & 
    & $^\dagger$$10^{0}$ & $^\dagger$$10^{-0.25}$ & 0.7 & 2 & 0.413 & 
    & $10^{0.25}$ & $^\star$$10^{0}$ & 0 & 1 & 0.599 \\ 
    $10^{0.25}$ & $10^{-0.75}$ & 0.7 & 2 & 0.178 & 
    & $^\dagger$$10^{0}$ & $^\dagger$$10^{-0.25}$ & 0.9 & 2 & 0.091 & 
    & $10^{0.25}$ & $^\star$$10^{0}$ & 0.7 & 1 & 0.103 \\ 
    $10^{0.25}$ & $10^{-0.75}$ & 0.9 & 2 & 0.058 & 
    & $^\dagger$$10^{0}$ & $^\dagger$$10^{-0.25}$ & 0.7 & 3 & 0.610 & 
    & $10^{0.25}$ & $^\star$$10^{0}$ & 0 & 3 & 1.000 \\ 
    \cdashline{7-11}
    $10^{0.25}$ & $10^{-0.75}$ & 0.3 & 3 & 0.921 & 
    & \multicolumn{5}{c}{\emph{atmosphere-hosting impactors}} & 
    & $10^{0.25}$ & $^\star$$10^{0}$ & 0.7 & 3 & 0.671 \\ 
    $10^{0.25}$ & $10^{-0.75}$ & 0.5 & 3 & 0.691 & 
    & $10^{-0.5}$ & $^\star$$10^{-0.5}$ & 0 & 1 & 0.475 & 
    & $10^{0.25}$ & $^\star$$10^{0.25}$ & 0 & 1 & 0.675 \\
		\cline{0-4} \cline{7-11} \cline{13-17}
	\end{tabular} \end{center}
	\caption{Table~\ref{tab:X_results}, continued.
    The first dashed line precedes the simulations 
    using ANEOS forsterite mantles, 
    indicated by a $^\dagger$ next to their masses.
    The second dashed line 
    indicates the start of the additional first-suite scenarios 
    with atmosphere-hosting impactors,
    indicated by a $^\star$ next to their mass.
    This includes scenarios where these impactors are treated as the targets 
    to give impactor:target mass ratios greater than one.
    Continued in Table~\ref{tab:X_results_3}.
    \label{tab:X_results_2}}
  \vspace{-1em}
\end{table*}

\begin{table*}[t]
  \small
	\begin{center} 
  \begin{tabular}{lllcc r lllcc r lllcc}
    \cline{0-4} \cline{7-11} \cline{13-17}
    $M_{\rm t}$ & $M_{\rm i}$ & $b$ & $v_{\rm c}$ & $X$ & 
    & $M_{\rm t}$ & $M_{\rm i}$ & $b$ & $v_{\rm c}$ & $X$ & 
    & $M_{\rm t}$ & $M_{\rm i}$ & $b$ & $v_{\rm c}$ & $X$ \\
    $\left(M_\oplus\right)$ & $\left(M_\oplus\right)$ & & $\left(v_{\rm esc}\right)$ & &
    & $\left(M_\oplus\right)$ & $\left(M_\oplus\right)$ & & $\left(v_{\rm esc}\right)$ & &
    & $\left(M_\oplus\right)$ & $\left(M_\oplus\right)$ & & $\left(v_{\rm esc}\right)$ & \\
		\cline{0-4} \cline{7-11} \cline{13-17}
    $^\star$$10^{-0.5}$ & $10^{-0.25}$ & 0.7 & 1 & 0.172 & 
    & $^\star$$10^{-0.25}$ & $10^{0}$ & 0.7 & 1 & 0.173 & 
    & $^\star$$10^{-0.25}$ & $10^{0.25}$ & 0.7 & 1 & 0.183 \\ 
    $^\star$$10^{-0.5}$ & $10^{-0.25}$ & 0 & 3 & 1.000 & 
    & $^\star$$10^{-0.25}$ & $10^{0}$ & 0 & 3 & 1.000 & 
    & $^\star$$10^{-0.25}$ & $10^{0.25}$ & 0 & 3 & 0.999 \\ 
    $^\star$$10^{-0.5}$ & $10^{-0.25}$ & 0.7 & 3 & 0.860 & 
    & $^\star$$10^{-0.25}$ & $10^{0}$ & 0.7 & 3 & 0.871 & 
    & $^\star$$10^{-0.25}$ & $10^{0.25}$ & 0.7 & 3 & 0.967 \\ 
    $^\star$$10^{-0.5}$ & $10^{0}$ & 0 & 1 & 0.340 & 
    & $^\star$$10^{-0.5}$ & $10^{0.25}$ & 0 & 1 & 0.725 & 
    & $^\star$$10^{0}$ & $10^{0.25}$ & 0 & 1 & 0.604 \\ 
    $^\star$$10^{-0.5}$ & $10^{0}$ & 0.7 & 1 & 0.194 & 
    & $^\star$$10^{-0.5}$ & $10^{0.25}$ & 0.7 & 1 & 0.197 & 
    & $^\star$$10^{0}$ & $10^{0.25}$ & 0.7 & 1 & 0.158 \\ 
    $^\star$$10^{-0.5}$ & $10^{0}$ & 0 & 3 & 1.000 & 
    & $^\star$$10^{-0.5}$ & $10^{0.25}$ & 0 & 3 & 0.998 & 
    & $^\star$$10^{0}$ & $10^{0.25}$ & 0 & 3 & 1.000 \\ 
    $^\star$$10^{-0.5}$ & $10^{0}$ & 0.7 & 3 & 0.978 & 
    & $^\star$$10^{-0.5}$ & $10^{0.25}$ & 0.7 & 3 & 1.000 & 
    & $^\star$$10^{0}$ & $10^{0.25}$ & 0.7 & 3 & 0.882 \\ 
    $^\star$$10^{-0.25}$ & $10^{0}$ & 0 & 1 & 0.615 & 
    & $^\star$$10^{-0.25}$ & $10^{0.25}$ & 0 & 1 & 0.451 & 
    & \\
		\cline{0-4} \cline{7-11} \cline{13-17}
	\end{tabular}
  \begin{tabular}{lccccc r lccccc}
    \multicolumn{6}{c}{\emph{third suite}} \\
    \cline{0-5} \cline{8-13}
    $M^{\rm base}_{\rm i}$ & \multicolumn{2}{c}{Target} & \multicolumn{2}{c}{Impactor} & $X$ &
    & $M^{\rm base}_{\rm i}$ & \multicolumn{2}{c}{Target} & \multicolumn{2}{c}{Impactor} & $X$ \\
    $\left(M_\oplus\right)$ & Mat. & Same & Mat. & Same & & 
    & $\left(M_\oplus\right)$ & Mat. & Same & Mat. & Same & \\
		\cline{0-5} \cline{8-13}
    $10^{-1}$ &  &  & Iron & $M$ & 0.453 & 
    & $10^{-0.5}$ & Rock & $M$ & Iron & $R$ & 0.858 \\ 
    $10^{-1}$ &  &  & Rock & $M$ & 0.324 & 
    & $10^{-0.5}$ & Rock & $M$ & Rock & $R$ & 0.524 \\ 
    $10^{-1}$ & Iron & $M$ &  &  & 0.261 & 
    & $10^{-0.5}$ & Iron & $R$ &  &  & 0.288 \\ 
    $10^{-1}$ & Iron & $M$ & Iron & $M$ & 0.350 & 
    & $10^{-0.5}$ & Iron & $R$ & Iron & $M$ & 0.451 \\ 
    $10^{-1}$ & Iron & $M$ & Rock & $M$ & 0.256 & 
    & $10^{-0.5}$ & Iron & $R$ & Rock & $M$ & 0.296 \\ 
    $10^{-1}$ & Rock & $M$ &  &  & 0.346 & 
    & $10^{-0.5}$ & Iron & $R$ & Iron & $R$ & 0.649 \\ 
    $10^{-1}$ & Rock & $M$ & Iron & $M$ & 0.481 & 
    & $10^{-0.5}$ & Iron & $R$ & Rock & $R$ & 0.270 \\ 
    $10^{-1}$ & Rock & $M$ & Rock & $M$ & 0.352 & 
    & $10^{-0.5}$ & Rock & $R$ &  &  & 0.576 \\ 
    $10^{-1}$ & Iron & $R$ & Iron & $R$ & 0.393 & 
    & $10^{-0.5}$ & Rock & $R$ & Iron & $M$ & 0.674 \\ 
    $10^{-1}$ & Rock & $R$ & Rock & $R$ & 0.352 & 
    & $10^{-0.5}$ & Rock & $R$ & Rock & $M$ & 0.590 \\ 
    $10^{-0.5}$ &  &  & Iron & $M$ & 0.606 & 
    & $10^{-0.5}$ & Rock & $R$ & Iron & $R$ & 0.860 \\ 
    $10^{-0.5}$ &  &  & Rock & $M$ & 0.525 & 
    & $10^{-0.5}$ & Rock & $R$ & Rock & $R$ & 0.542 \\ 
    $10^{-0.5}$ &  &  & Iron & $R$ & 0.768 & 
    & $10^{0}$ &  &  & Iron & $M$ & 0.802 \\ 
    $10^{-0.5}$ &  &  & Rock & $R$ & 0.478 & 
    & $10^{0}$ &  &  & Rock & $M$ & 0.746 \\ 
    $10^{-0.5}$ & Iron & $M$ &  &  & 0.379 & 
    & $10^{0}$ & Iron & $M$ &  &  & 0.686 \\ 
    $10^{-0.5}$ & Iron & $M$ & Iron & $M$ & 0.553 & 
    & $10^{0}$ & Iron & $M$ & Iron & $M$ & 0.786 \\ 
    $10^{-0.5}$ & Iron & $M$ & Rock & $M$ & 0.363 & 
    & $10^{0}$ & Iron & $M$ & Rock & $M$ & 0.724 \\ 
    $10^{-0.5}$ & Iron & $M$ & Iron & $R$ & 0.747 & 
    & $10^{0}$ & Rock & $M$ &  &  & 0.824 \\ 
    $10^{-0.5}$ & Iron & $M$ & Rock & $R$ & 0.351 & 
    & $10^{0}$ & Rock & $M$ & Iron & $M$ & 0.884 \\ 
    $10^{-0.5}$ & Rock & $M$ &  &  & 0.564 & 
    & $10^{0}$ & Rock & $M$ & Rock & $M$ & 0.827 \\ 
    $10^{-0.5}$ & Rock & $M$ & Iron & $M$ & 0.662 & 
    & $10^{0}$ & Iron & $R$ & Iron & $R$ & 0.910 \\ 
    $10^{-0.5}$ & Rock & $M$ & Rock & $M$ & 0.578 & 
    & $10^{0}$ & Rock & $R$ & Rock & $R$ & 0.782 \\ 
		\cline{0-5} \cline{8-13}
  \end{tabular} 
  \end{center}
	\caption{Table~\ref{tab:X_results_2}, continued.
    The separate headings precede 
    the simulations in the third suite of different-density bodies.
    All of these scenarios are based on the $10^0$~$M_\oplus$ target
    with $b = 0.7$ and $v_{\rm c} = 3$~$v_{\rm esc}$.
    The remaining parameters are the base impactor mass,
    the material of each body and whether their mass or radius 
    was kept the same as the base version,
    or left blank for a standard body with both materials.
    \label{tab:X_results_3}}
  \vspace{-1em}
\end{table*}

\section{Approximate Interacting Mass} \label{sec:appx:scaling}

The fractional interacting mass,
which for any impact angle loosely accounts for the proportions
of the two bodies that interact, is given by 
\begin{equation}
  f_M \equiv 
    \dfrac{\rho_{\rm t} V^{\rm cap}_{\rm t} + \rho_{\rm i} V^{\rm cap}_{\rm i}}
    {\rho_{\rm t} V_{\rm t} + \rho_{\rm i} V_{\rm i}} \;,
  \label{eqn:M_iact}
\end{equation}
\\ 
where $V_{\rm t,\,i}$ are the total volumes of each body,
ignoring any atmosphere,
and $V^{\rm cap}_{\rm t,\,i}$ are the volumes of 
the target cap above the lowest point of the impactor at contact
and the impactor cap below the highest point of the target, respectively.
Both caps have height ${d = \left(R_{\rm t} + R_{\rm i}\right) (1 - b)}$, giving
$V^{\rm cap}_{\rm t,\,i} = \tfrac{\rm \pi}{3} d^2 \left(3 R_{\rm t,\,i} - d\right)$.

For equal bulk densities, this simplifies to 
the fractional interacting volume from \citet[][Appx.~B]{Kegerreis+2020}:
\begin{equation}
  f_V \equiv 
    \dfrac{V^{\rm cap}_{\rm t} + V^{\rm cap}_{\rm i}}
    {V_{\rm t} + V_{\rm i}} = \tfrac{1}{4}
    \dfrac{\left(R_{\rm t} + R_{\rm i}\right)^3}{R_{\rm t}^3 + R_{\rm i}^3} 
    (1-b)^2 (1+2b) \;.
  \label{eqn:V_iact}
\end{equation}
For the collisions in this study,
$f_M$ only differs from $f_V$ by a median relative change of 2.5\%.


\bibliography{gihr.bib}{}
\bibliographystyle{aasjournal}



\end{document}